\documentclass[aps,prb,amsmath,amsfonts,amssymb,superscriptaddress,twocolumn,showpacs]{revtex4}
\renewcommand{\vec}[1]{\mathbf{#1}}
\usepackage[]{graphicx}
\usepackage{wasysym}
\usepackage{setspace}
\usepackage{color}
\usepackage{nicefrac}
\usepackage{times}

\renewcommand{\Re}{\operatorname{Re}}
\renewcommand{\Im}{\operatorname{Im}}

\newcommand{\figref}[1]{Fig.~\ref{fig:#1}}
\newcommand{\Figref}[1]{Figure~\ref{fig:#1}}

\renewcommand{\eqref}[1]{Eq.~(\ref{eq:#1})}

\newcommand{\eqreftwo}[2]{Eqs.~(\ref{eq:#1}) and (\ref{eq:#2})}
\newcommand{\Eqreftwo}[2]{Equations~(\ref{eq:#1}) and (\ref{eq:#2})}
\newcommand{\eqrefrange}[2]{Eqs.~(\ref{eq:#1})--(\ref{eq:#2})}
\newcommand{\Eqref}[1]{Equation~(\ref{eq:#1})}
\newcommand{\citeasnoun}[1]{Ref.~\onlinecite{#1}}

\newcommand{\appref}[1]{Appendix~\ref{app:#1}}

\newcommand{\Teff}{T_\mathrm{eff}} 
\newcommand{\Td}{T_\mathrm{d}}
\newcommand{\Te}{T_\mathrm{e}}
\newcommand{\TeffNL}{T^\mathrm{NL}_\mathrm{eff}}
\newcommand{\dw}{\delta\omega} 
\newcommand{\kB}{k_\mathrm{B}}
\newcommand{\da}{\delta a}

\def\a{s}
\def\b{s}
\newcommand{\add}[1]{\if\a\b{{\color{black} #1}}\else{#1}\fi}
\newcommand{\comm}[1]{\if\a\b{{\color{blue}\{\small \sc #1\}}}\else{}\fi}
\newcommand{\del}[1]{{\if\a\b{{\color{magenta}[[#1]]}}\else{}\fi}}

\begin{document}

\title{Radiative heat transfer in nonlinear Kerr media}

\author{Chinmay Khandekar}
\affiliation{Department of Electrical Engineering, Princeton University, Princeton, NJ 08540}
\author{Adi Pick}
\affiliation{Department of Physics, Harvard University, Cambridge, MA 02138}
\author{Steven G. Johnson}
\affiliation{Department of Mathematics, Massachusetts Institute of Technology, Cambridge, MA 02139}
\author{Alejandro W. Rodriguez}
\affiliation{Department of Electrical Engineering, Princeton University, Princeton, NJ 08540}

\begin{abstract}
  We obtain a fluctuation--dissipation theorem describing thermal
  electromagnetic fluctuation effects in nonlinear media that we
  exploit in conjunction with a stochastic Langevin framework to study
  thermal radiation from Kerr ($\chi^{(3)}$) photonic cavities coupled
  to external environments at and out of equilibrium. We show that
  that in addition to thermal broadening due to two-photon absorption,
  the emissivity of such cavities can exhibit asymmetric,
  non-Lorentzian lineshapes due to self-phase modulation. When the
  local temperature of the cavity is larger than that of the external
  bath, we find that the heat transfer into the bath exceeds the
  radiation from a corresponding linear black body at the same local
  temperature. We predict that these temperature-tunable thermal
  processes can be observed in practical, nanophotonic cavities
  operating at relatively small temperatures.
\end{abstract} 

\date{\today}

\maketitle 

%\tableofcontents

%including thermal rectification~[refs], nano-imaging~[refs],
%nanoscale cooling~[ref], and thermophotovoltaic power
%generation~[ref].

\section{Introduction}

The radiative properties of bodies play a fundamental role on the
physics of many naturally occurring processes and emerging
nanotechnologies~\cite{satoshi09,lenert14}. Central to the theoretical
understanding of these electromagnetic fluctuation effects is the
fluctuation-dissipation theorem of electromagnetic fields, developed
decades ago by Rytov and
others~\cite{Rytov89,PolderVanHove71,Eckhardt84} in order to describe
radiative transport in macroscopic media. The same formalism has been
recently employed in combination with new theoretical
techniques~\cite{ReidRo12:review,Otey14:review} to demonstrate strong
modifications of the thermal properties of nanostructured bodies,
including designable selective emitters~\cite{Laroche06:prl} and
greater than blackbody heat transport between bodies in the
near-field~\cite{BasuZhang09}. To date, these studies have focused
primarily on linear media, where emission depends only on the linear
response functions of the underlying materials. A cubic ($\chi^{(3)}$)
nonlinearity, however, can convert light from one frequency to another
or alter the dissipation rate~\cite{Boyd92} and hence the fluctuation
statistics. We show that these phenomena lead to a variety of
interesting effects in nonlinear radiators, such as lineshape
alterations, temperature-dependent emission, and even radiation
exceeding the black-body limit in nonequilibrium systems.

In this work, we obtain a nonlinear fluctuation--dissipation theorem
(FDT) that describes radiative thermal effects in nonlinear
$\chi^{(3)}$ media, extending previous work on nonlinear
oscillators~\cite{dykman75}. Since nonlinear optical effects are
generally weak in bulk materials, we focus on nanostructured resonant
systems with strong effective nonlinear
interactions~\cite{Boyd92,Bravo-Abad07}. Such systems are susceptible
to universal descriptions based on the coupled-mode theory
framework~\cite{Haus84,Hashemi09}, which we exploit to investigate the
ways in which nonlinearities can enable interesting/designable
radiative effects. In particular, we show that self-phase modulation
(SPM) and two-photon absorption (TPA) effects lead to strong
modifications of their emissivity, including thermal broadening and
non-Lorentzian, asymmetric lineshapes. These nonlinear effects pave
the way for additional material tunability, including designable,
temperature-dependent selective emitters and absorbers. We also
consider nonequilibrium situations and show that TPA results in
selective heat transfer exceeding the black-body limit, a phenomenon
that has only been observed in situations involving multiple bodies in
the near-field~\cite{BasuZhang09}. Finally, we show that recently
proposed, wavelength-scale cavities with ultra-large dimensionless
lifetimes $Q \lesssim 10^8$ and small mode volumes $V \sim
(\lambda/n)^3$ can be designed to display these strongly nonlinear
effects at infrared wavelengths and near room temperature.

Fluctuation--dissipation relations in nonlinear media have been a
subject of much interest in recent decades, starting with the early
work of Bernard and Callen~\cite{berncallen},
Stratonovich~\cite{stratonovichBook93}, and
Klimontovich~\cite{klim}. The effects of nonlinearities of both
conservative and dissipative nature on the Brownian motion of resonant
systems have been studied in the context of Van der Pol
oscillators~\cite{klim}, optomechanical systems~\cite{Kippenberg07},
and mechanical Duffing oscillators~\cite{dykman75,Zaitsev12}. Despite
the relatively large body of work involving noise in nonlinear
systems, the role and consequences of nonlinear damping in mechanical
oscillators have only recently begun to be
explored~\cite{Zaitsev12,Moser12}, and there remains much to be known
about the underlying physical mechanisms in such systems. The effects
of nonlinear noise are also non-negligible and of great importance in
a variety of applications, e.g. MEMS sensors~\cite{Cleland02},
frequency stabilization~\cite{Antonio12}, frequency
mixing~\cite{HoChan07}, and filtering~\cite{Buks06}. While there is
increased interest in studying nonlinear effects in micro and
nano-mechanical oscillators, studies of nonlinear effects on thermal
radiation remain scarce and are largely restricted to driven systems
with conservative nonlinearities e.g. resonators based on RF-driven
Josephson junctions~\cite{Andre12} or optomechanical
oscillators~\cite{Kippenberg:07}. (The situation is different in the
quantum regime, where the effects of SPM on the tunneling rate and
quantum statistics of photons have been well
studied.)~\cite{Walls80,dykman84} Following an approach analogous to
the treatment of nonlinear friction in mechanical
oscillators~\cite{dykman75}, we extend previous work on Duffing
oscillators to the case of nonlinear photonic cavities coupled to
external baths/channels, a situation of direct relevance to
current-generation experiments on radiative thermal transport in
photonic media~\cite{Luo04:thermal}. Interestingly, we find that
effects arising from the interference of radiation reflected and
emitted from the cavity into the external bath are crucial in order to
observe thermal radiation enhancements in realistic situations, such
as in cases where the external-bath temperature is at or near room
temperature. We believe that these photonic systems not only offer new
opportunities for understanding the role of nonlinear damping on
fluctuations, but also greatly extend the functionality and tunability
of devices based on thermal radiation. As we argue below, while these
effects require very strong optical nonlinearities, the increasing
accessibility of ultra-high $Q$ resonators with small modal
volumes~\cite{Vahala03,Bravo-Abad07,Qimin11,Painter10,Johnson01:cavities},
such as the nanobeam cavity explored below, offers hope that they may
soon be within the reach of experiments.

\section{Langevin Framework}

We begin by introducing the Langevin equations of motion of a
single-mode nonlinear $\chi^{(3)}$ cavity coupled to an external bath
(a single output channel) and an internal reservoir (a lossy
channel). As described in~\citeasnoun{Rodriguez07:OE}, the
coupled-mode equations for the field amplitude are given by:
\begin{eqnarray}
\label{eq:cavity1}
  \frac{da}{dt} &=& [i(\omega_{0} - \alpha|a|^{2})-\gamma]a +
  \sqrt{2\gamma_{e}} s_{+} + D \xi, \\ s_{-} &=& -s_{+} +
  \sqrt{2\gamma_e} a,
\label{eq:cavity2}
\end{eqnarray} 
where $|a|^{2}$ is the mode energy, $|s_\pm|^2$ are the input/output
power from/to the external bath (e.g. a waveguide), and $\omega_0$ and
$\gamma=\gamma_e+\gamma_d$ are the frequency and linear decay rate of
the mode. The linear decay channels include linear absorption from
coupling to phonons or other dissipative degrees of freedom
($\gamma_d$) as well as decay into the external environment
($\gamma_e$). The real and imaginary parts of the nonlinear
coefficient $\alpha$ are given by the overlap integral $\alpha =
\frac{3}{4}\omega_0 \int \varepsilon_0 \chi^{(3)} |\vec{E}|^4 / (\int
\varepsilon |\vec{E}|^2)^2$ of the linear cavity fields $\vec{E}$ and
lead to SPM and TPA, respectively.~\cite{Rodriguez07:OE} In addition
to radiation coming from the external bath $\sim s_+$, \eqref{cavity1}
includes a stochastic Langevin source $D \xi(t)$ given by the product
of a normalized ``diffusion coefficient'' $D$, relating amplitude
fluctuations to dissipation from the internal (phonon) reservoir, and
a time-dependent stochastic process $\xi(t)$ whose form and properties
can be derived from very general statistical
considerations~\cite{vankampen,stratonovichBook93,Zwanzig01:book}. For
linear systems ($\alpha=0$), the stochastic terms are uncorrelated
white-noise sources (assuming a narrow bandwidth $\gamma \ll
\omega_0$) satisfying:
\begin{eqnarray}
\label{eq:sfdt}
  \langle s_{+}^{*}(t)s_{+}(t') \rangle &=& \kB \Te \delta(t-t'),
  \\ \langle \xi^{*}(t)\xi(t') \rangle &=& \kB \Td \delta(t-t'),
  \\ D(\gamma_d) &=& \sqrt{2\gamma_d},
\label{eq:xifdt}
\end{eqnarray}
where $\langle \ldots \rangle$ is a thermodynamic ensemble average,
and $T_{d}$ and $\Te$ are the local temperatures of the internal and
external baths, respectively.

The presence of nonlinear dissipation $\sim \Im \alpha |a|^2$ means
that $D$ must also depend on $a$ and $\Im\alpha$.~\cite{vankampen}
Note that $\Re\alpha$ does not play any role in nonlinear
dissipation. This intuitive result also follows from a microscopic
Hamiltonian approach where $\Re\alpha$ appears in the isolated system
Hamiltonian as the quartic nonlinearity term while $\Im\alpha$
represents system-heat bath nonlinear coupling~\cite{dykman84}. As a
result, the diffusion coefficient $D$ which captures the cavity--bath
interaction in \eqref{cavity1} does not depend on
$\Re\alpha$. (Interestingly, in the case of a driven quantum
oscillator, the real part of $\chi^{(3)}$ affects the tunneling rate
between states and hence the fluctuation statistics~\cite{Walls80}.)
Such a nonlinear FDT can be obtained from very general statistical
considerations such as energy
equipartition~\cite{vankampen,stratonovichBook93,Zwanzig01:book},
derived under the assumption that the system is at equilibrium,
i.e. $T=\Te=\Td$. As described in \appref{FP}, one can apply a
standard procedure to transform the stochastic ODE [\eqref{cavity1}]
into a Fokker--Planck equation for the probability distribution
$P(a,a^*)$,~\cite{MossBook07} which in our case is given by:
\begin{multline}
  \frac{dP(a,a^*)}{dt} =-\frac{\partial}{\partial a}K_{a}P-\frac{\partial}{\partial a^*}K_{a^*}P + \frac{1}{2} \frac{\partial^2}{\partial a \partial a^*} K_{a a^*} P
\label{eq:dpdt}
\end{multline}
%\begin{equation}
%\frac{dP(a,a^*)}{dt} = -\frac{\partial}{\partial a_{\alpha}}K_{a_{\alpha}}P
%  + \frac{1}{2} \frac{\partial^2}{\partial a_{\alpha}\partial a_{\beta}} 
%  K_{a_{\alpha}a_{\beta}}P,
%\label{eq:dpdt}
%\end{equation}
with Fokker--Planck coefficients,
\begin{align*}
  K_a &= [i(\omega_0-\Re\alpha|a|^2)-(\gamma -\Im\alpha|a|^2)]a +
  \lambda D\frac{\partial D}{\partial a^*} \\
  K_{a^{*}} &= [-i(\omega_0-\Re\alpha|a|^2)-(\gamma -\Im\alpha|a|^2)]a^* +
  \lambda D\frac{\partial D}{\partial a} \\
  K_{aa^*}&=K_{a^*a}=(2\gamma_e+D^2),\hspace{10pt} K_{aa}=K_{a^*a^*}=0 
\end{align*}
The precise meaning of these coefficients depends on the integration
rule used to describe the stochastic ODE. Here,
$\lambda=0,\frac{1}{2},1$ correspond to Ito, Stratonovich and kinetic
interpretations of stochastic calculus, respectively. The parameters
$\lambda$ and $D$ must of course ensure that the statistical
properties of the system are consistent with the laws of
thermodynamics.

Based on the standard working hypothesis of statistical mechanics, the
equilibrium state is described by the Maxwell-Boltzmann distribution
$e^{-U/\kB T}$, where $U$ is the energy in the cavity field. Since the
nonlinearity is considered perturbatively in the derivation of the
coupled mode equations,~\cite{Haus84:coupled} nonlinear contributions
to the cavity energy $|a|^2$ can be safely neglected.~\footnote{To
  first order in $\chi^{(3)}$, the only impact of the nonlinearity is
  to shift the cavity frequency slightly which does not contribute to
  changes in the cavity energy~\cite{JoannopoulosJo08-book}.} More
generally however, for an undriven oscillator, the nonlinear
contribution to the oscillator energy can be ignored if the shift in
the frequency is much smaller than the cavity
eigenfrequency~\cite{dykman84}, which is the case for the passive,
nonlinear cavity considered here. This well-known result can be
verified by solving \eqref{dpdt} in the absence of nonlinear
dissipation ($\Im\alpha=0$), in which case the fluctuation statistics
remain unchanged and one finds that terms $\sim \Re\alpha$ do not
affect the steady-state probability
distribution~\cite{vankampen,Kardar07}.  It follows that the
equilibrium state of the system is described by the Gibbs distribution
$e^{-|a|^2/\kB T}$ and the stochastic equations must be interpreted
according to the kinetic rule $\lambda=1$, provided that the diffusion
coefficient $D$ have the form:
\begin{equation}
  D(a,a^*)=\sqrt{2(\gamma_d-\Im\alpha|a|^2)}
\label{eq:Dnl}
\end{equation}
Hence, the only modification to the diffusion coefficient is the
addition of the nonlinear dissipation rate $\Im \alpha |a|^2$. Note
that the particular form or choice of multiplicative noise will
determine the corresponding stochastic interpretation, and vice
versa. For instance, it is also possible to choose the internal noise
in \eqref{cavity1} to be of the form $\sqrt{2\gamma_d} \xi_1 +
D(a,a^*) \xi_2$, where $\xi_{1,2}$ are independent Gaussian noise
sources, provided that $D=\sqrt{-2\Im\alpha}a^*$ and that the
stochastic ODE is interpreted according to the Stratonovich rule
$\lambda=\frac{1}{2}$.~\cite{dykman84} Our choice of interpretation
here is chosen purely for convenience.

\section{Thermal Radiation}

\Eqreftwo{cavity1}{Dnl} can be solved to obtain both the equilibrium
and nonequilibrium behavior of the system. Since they do not admit
closed-form analytical solutions, we instead solve the stochastic ODE
numerically using the Euler--Maruyama method~\cite{Higham01},
involving a simple forward-difference discretization which for the
kinetic calculus results in additional terms compared to an Ito
discretization~\cite{MossBook07}. To lowest order in the
discretization,
\begin{align*}
  \Delta a = [&i(\omega_{0} - \alpha|a|^{2}) - \gamma] a \Delta t + D
  \Delta W_\xi \\ &+ \frac{\partial D}{\partial a}\Delta a \Delta W_\xi
  + \frac{\partial D}{\partial a^{*}} \Delta a^{*} \Delta W_\xi +
  \sqrt{2\gamma_e} \Delta W_{s+},
\end{align*}
where $\Delta a \equiv a(t+\Delta t)-a(t)$ and $\Delta W_{f} \equiv
W_f(t+\Delta t)-W_f(t) = f \Delta t$ is a Wiener
process~\cite{Higham01} corresponding to the white-noise stochastic
signal $f \in \{\xi,s_+\}$. It follows that to first order in $\Delta
t$, the discretized ODE is given by:
\begin{align}
  \Delta a = [i(&\omega_0 - \alpha |a|^{2}) + \Im \alpha |\xi|^2-
  \gamma] a \Delta t \nonumber \\ &+ 2\sqrt{\gamma_d - \Im \alpha
    |a|^2} \Delta W_\xi + \sqrt{2\gamma_e} \Delta W_{s+},
\label{eq:Da}
\end{align}
where the additional discretization term $\sim \Im \alpha |\xi|^2$
arises in the kinetic and not the Ito calculus.

\begin{figure}[t!]
\includegraphics[width=0.95\linewidth]{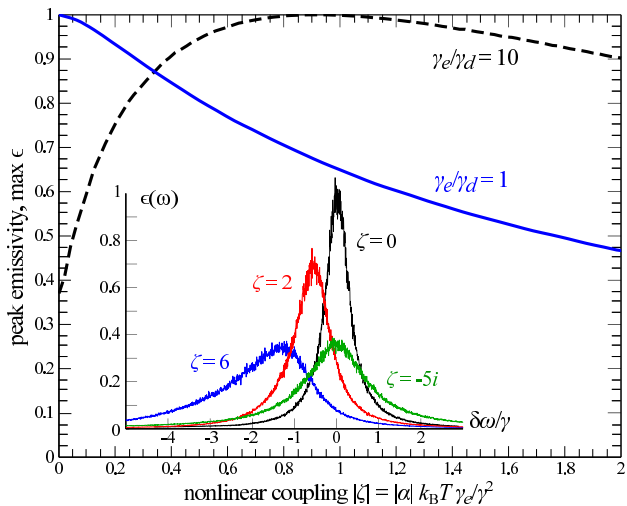}
\caption{Peak emissivity $\epsilon_\mathrm{max}$ of a cavity coupled
  to an external bath, both at temperature $T$, as a function of
  nonlinear coupling $|\zeta| = |\alpha| \kB T \gamma_e/\gamma^2$, for
  different ratios of the linear dissipation $\gamma_e$ and external
  coupling $\gamma_d$ rates. Inset shows the emissivity
  $\epsilon(\omega)$ for $\gamma_e=\gamma_d$, corresponding to a
  cavity with perfect linear emissivity, for multiple values of
  $\zeta$, illustrating the effects of SPM (red/blue) and TPA (green)
  on the spectrum.}
\label{fig:fig1}
\end{figure}

\emph{Equilibrium}---In what follows, we demonstrate numerically that
the system described by \eqreftwo{cavity1}{Dnl} thermalizes and
satisfies all of the properties of an equilibrium thermodynamic
system, including equipartition and detailed balance, but that
nonlinearities lead to strong modifications of the emissivity of the
cavity.  We consider the equilibrium situation $T\equiv \Td=\Te$, in
which case $\langle |s_{+}|^2 \rangle = \langle |\xi|^2 \rangle = \kB
T$. To begin with, we motivate our numerical results by performing a
simple mean-field approximation known as statistical
lineariation~\cite{Gieseler13}, which captures basic features but
ignores correlation effects stemming from
nonlinearities. Specifically, making the substitution $|a(t)|^2 \to
\langle |a(t)|^2 \rangle = \kB T$ in \eqref{cavity1}, and solving for
the steady-state, linear response of the system, we obtain the
emissivity of the cavity $\epsilon(\omega) \equiv 2\gamma_e \langle
|a(\omega)|^2 \rangle / \kB T$, defined as the emitted power into the
external bath normalized by $\kB T$ in the limit $s_+ \to 0$. In
particular, we find:
\begin{align}
  \epsilon(\omega) = \frac{4\gamma_e(\gamma_d-\Im \alpha \kB T)}{\dw^2_T
    +(\gamma-\Im\alpha \kB T)^2} \leq 1,
\label{eq:awT}
\end{align}
where $\dw_T \equiv \omega-\omega_0 + \Re \alpha \kB T$ and $\epsilon
\leq 1$ as expected from Kirchoff's law~\cite{Rytov89,Eckhardt84}.

\Eqref{awT} can be integrated to verify the self-consistency condition
$\langle |a(t)|^2 \rangle = \int \frac{d\omega}{2\pi}\, \langle
|a(\omega)|^2 \rangle = \kB T$, as required by equipartition. It can
also be combined with \eqref{cavity2} to show that detailed balance
$\langle|s_{-}(\omega)|^2\rangle = \langle|s_{+}(\omega)|^2\rangle$ is
satisfied, i.e. there is no net transfer of heat from the cavity to
the external bath and vice versa. More interestingly, the presence of
$\alpha$ leads to a temperature-dependent change in the frequency and
bandwidth of the cavity proportional to $\Re \alpha$ and $\Im \alpha$,
respectively. These properties are validated by a full solution of the
ODE, as illustrated on the inset of~\figref{fig1}, which shows the
numerically computed emissivity $\epsilon(\omega)$ for a few values of
the dimensionless nonlinear coupling $\zeta \equiv \alpha \kB T
\gamma_e / \gamma^2$. Although \eqref{awT} yields good agreement with
our numerical results for small $|\zeta| \lesssim 0.5$, at larger
temperatures correlation effects become relevant and statistical
linearization is no longer able to describe (even qualitatively) the
spectral features. For instance, in the absence of TPA and for large
$\zeta$ (such as $\zeta=6$ in \figref{fig1}), SPM leads to
asymmetrical broadening of the spectrum: broadening is most pronounced
along the direction of the frequency shift, as determined by the sign
of $\Re \alpha$. This effect is known as ``frequency straddling'', has
been predicted in the context of Duffing mechanical
oscillators,~\cite{dykman92,weiss87,MossBook07} and arises due to
frequency mixing within the cavity bandwidth, as captured by the
perturbative expansion of the emissivity in powers of $\alpha$ in
\eqref{aneq}. In particular, at equilibrium one finds that the
first-order correction to the emissivity $\sim -(\Re\alpha \kB T)$,
and so SPM enhances and reduces thermal contributions from red and
blue-detuned frequencies, or vice versa depending on the sign of
$\Re\alpha$. Intuitively, the density of states within the cavity
favors frequency conversion away from the resonance. Hence, photons on
the red side of the resonance will experience larger frequency shifts
than those on the blue side for $\Re \alpha > 0$ (red shifting), and
vice versa for $\Re \alpha < 0$ (blue shifting). Note that
equipartition $\langle |a|^2 \rangle =k_BT$ and detailed balance
$\langle |s_{+}(\omega)|^2\rangle = \langle |s_{-}(\omega)|^2 \rangle
$ are satisfied even in the presence of strong correlations.

The above SPM and TPA effects pave the way for designing
temperature-tunable thermal emissivities. For instance, it is well
known that in a linear system, a cavity can become a perfect
emitter/absorber when the emission and dissipation rates are equal,
i.e. $\gamma_e=\gamma_d$.~\cite{JoannopoulosJo08-book} It follows from
\eqref{awT} that in the nonlinear case there is a modified
rate-matching condition whereby $\epsilon=1$ is achieved only at the
critical temperature $T_c$ where $\gamma_e = \gamma_d - \Im \alpha
(\kB T_c)$.  Hence, a system designed to have $\gamma_e > \gamma_d$
(since $\Im \alpha < 0$ in any passive system~\cite{Boyd92}) at room
temperature will become a perfect emitter at $T_c \gtrsim 300$K. To
illustrate this phenomenon, \figref{fig1} shows the variation of the
peak emissivity of the cavity, $\epsilon_\mathrm{max}$, by tuning the
effective nonlinearity $|\zeta|$ for multiple values of
$\gamma_e/\gamma_d$.

\emph{Nonequilibrium}---We now consider nonequilibrium conditions and
demonstrate that TPA can lead to thermal radiation exceeding the
black-body limit. Assuming local equilibrium conditions, $\langle
|s_{+}|^{2} \rangle =\kB T_{e}$, $\langle |\xi|^{2}\rangle = \kB
T_{d}$, the heat transfer between the cavity and external bath is
given by:
\begin{align}
  H &= \int_{-\infty}^\infty \frac{d\omega}{2\pi} \left(\langle
  |s_{-}(\omega)|^2 \rangle - \langle |s_{+}(\omega)|^2 \rangle\right)
  \nonumber \\ &= \int_{-\infty}^\infty \frac{d\omega}{2\pi}\, \Phi(\omega) \kB
  \Delta T
\end{align}
where $\Delta T \equiv \Td-\Te$ and $\Phi(\omega)$ is known as the
spectral transfer function~\cite{Otey14:review}, or the heat exchange
between the two systems compared to two black bodies. (The transfer
function of a black body $\Phi_\mathrm{BB}(\omega) = 1$ at all frequencies.)

To begin with, we consider a perturbative expansion of \eqref{cavity1}
in powers of $\alpha$, described in \appref{pert}, which we
find to be accurate to within a few percent up to $|\zeta| \approx
0.5$. In this case we find that statistical linearization does not
even qualitatively describe the behavior of the system at small
$\alpha$. To linear order in $\alpha$, perturbation theory leads to
the following expressions for the energy and output-power spectra:
%\begin{align}
%  \label{eq:aneq}
%  \hspace{-0.1in}\langle |a(\omega)|^{2} \rangle &= \frac{2\gamma \kB
%    \Teff}{\dw^{2}+\gamma^{2}} - \frac{4\dw \gamma \Re\alpha
%    (\kB\Teff)^2}{(\dw^2+\gamma^2)^2} \nonumber \\ &- \frac{2\Im\alpha
%    (\kB^2 \Teff)}{(\dw^2+\gamma^2)} \left[\Td +
%    \frac{2\gamma^2(\Td-2\Teff)}{\dw^2+\gamma^2}\right]
%  \\ \Phi(\omega) &= \frac{4\gamma_e\gamma_d}{\dw^{2}+\gamma^{2}} -
%  \frac{8 \dw \gamma_e \gamma_d \Re\alpha (\kB \Teff)}{(\dw^2 +
%    \gamma^2)^2} \nonumber \\ &- \frac{1}{\Delta T}\frac{4\gamma_e
%    \Im\alpha \kB }{(\dw^{2}+\gamma^2)} \Big[\Teff T_{d} \nonumber
%    \\ &\left.+ \frac{[2\gamma^{2}\Teff +
%        (\dw^2-\gamma^2)\Te](T_{d}-2\Teff)}{\dw^{2}+\gamma^{2}}
%    \right],
%  \label{eq:Phi}
%\end{align}
\begin{widetext}
\begin{align}
\label{eq:aneq}
  \langle |a(\omega)|^{2} \rangle &= \frac{2\gamma \kB
    \Teff}{\dw^{2}+\gamma^{2}} - \frac{4\dw \gamma \Re\alpha
    (\kB\Teff)^2}{(\dw^2+\gamma^2)^2} + \frac{2\Im\alpha (\kB^2
    \Teff)}{(\dw^2+\gamma^2)} \left[\Td +
    \frac{2\gamma^2(\Td-2\Teff)}{\dw^2+\gamma^2}\right]
  \\ \Phi(\omega) &= \frac{4\gamma_e\gamma_d}{\dw^{2}+\gamma^{2}} -
  \frac{8 \dw \gamma_e \gamma_d \Re\alpha (\kB \Teff)}{(\dw^2 +
    \gamma^2)^2} - \frac{1}{\Delta T}\frac{4\gamma_e \Im\alpha \kB
  }{(\dw^{2}+\gamma^2)}\left[\Teff T_{d} + \frac{[2\gamma^{2}\Teff +
        (\dw^2-\gamma^2)\Te](T_{d}-2\Teff)}{\dw^{2}+\gamma^{2}}
    \right],
\label{eq:Phi}
\end{align}
\end{widetext}
where $\dw \equiv \omega-\omega_0$ and $\Teff = \frac{\gamma_e \Te +
  \gamma_d \Td}{\gamma}$ is the effective temperature $\langle
|a(t)|^{2}\rangle/\kB$ of the cavity in the linear regime. At finite
$\alpha$, the effective temperature is given by:
\begin{align}
  \TeffNL = \Teff - \frac{2\Im\alpha \kB \Teff}{\gamma}\left(\Td -
  \Teff\right),
\label{eq:Tnl}
\end{align}
which reduces to $\Teff$ in the absence of nonlinearities and at
equilibrium. Furthermore, one can also show that in the linear regime,
$\Phi \leq \Phi_\mathrm{BB}$ and reaches its maximum at the resonance
frequency when $\gamma_e=\gamma_d$. For finite $\Im \alpha \neq 0$, we
find that $\TeffNL > \Teff$ irrespective of system parameters and that
under certain conditions $\Phi$ increases above one. Thus, one arrives
at the result that out of equilibrium, the rate at which energy is
drawn from the phonon bath can be larger than the rate at which energy
radiates from the cavity, causing the effective temperature and
overall heat transfer to increase above its linear value, a phenomenon
associated with the presence of excess heat.~\cite{Oono98} Note that
$\TeffNL$ is not affected by $\Re \alpha$ to first order since the
perturbation is odd in $\dw$ and therefore integrates to zero. One can
show that the peak transfer can increase above one whenever
$\frac{d\Phi(\omega_0)}{d(-\Im\alpha)} \left[\Teff (3 T_{d} + 2\Te
  -4\Teff)-T_{e}T_{d}\right] > 0$ is satisfied, which occurs for
instance when $\Td \gg \Te$, in which case
$\frac{d\Phi(\omega_0)}{d(-\Im\alpha)} > 0$. Thus, when a linear
system has nearly perfect emissivity, any small amount of TPA can push
its radiation above the black-body limit. For example, the peak
emissivity of a system with $\gamma_e=\gamma_d$, $\Te=0$, and subject
to TPA, is given from \eqref{Tnl} by $\eta \equiv
\frac{\Phi(\omega_0)}{\Phi_\mathrm{BB}} = 1 - \frac{\Im \alpha (\kB
  \Td)}{2\gamma}$, which increases above one with increasing $-\Im
\alpha$.

\begin{figure}[t!]
\includegraphics[width=0.95\linewidth]{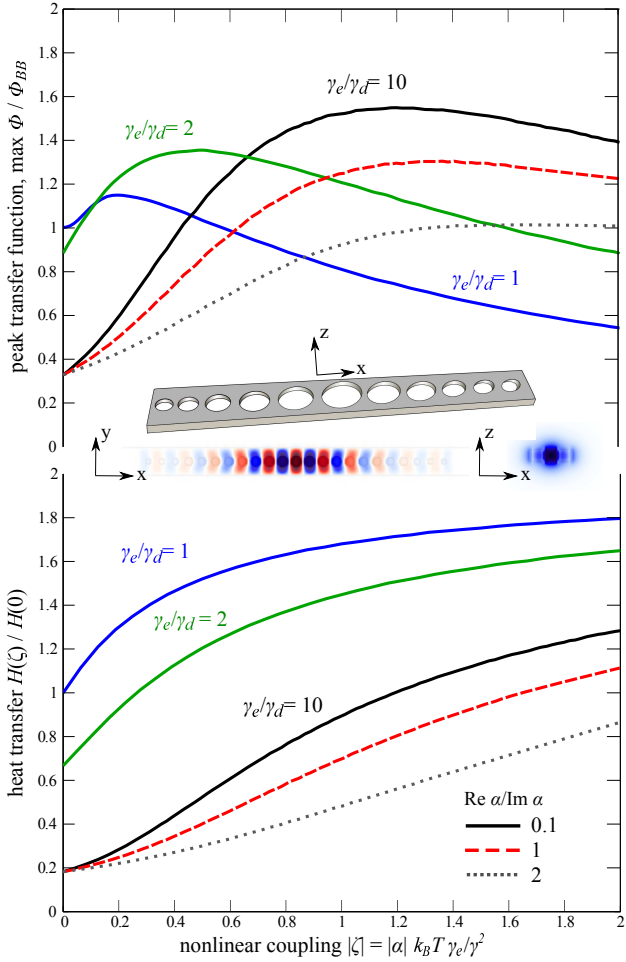}
\caption{Peak (on-resonance) spectral transfer function
  $\Phi_\mathrm{max} \equiv \Phi(\omega_0)$ normalized by the black
  body $\Phi_\mathrm{BB}$ (top), and net heat transfer $H(\zeta)$
  normalized by $H_{\max}(0)$ (bottom), as a function of nonlinear
  coupling $|\zeta| = |\alpha| \kB T \gamma_e/\gamma^2$, for a system
  consisting of a cavity at temperature $\Td$ coupled to an external
  bath at $\Te = 0$, for multiple configurations of
  $\gamma_e/\gamma_d$ and $\Re \alpha/\Im \alpha$. Inset shows a
  cavity design supporting a mode at $\lambda\approx
  2.09\mu\mathrm{m}$ with lifetime $Q \approx 10^8$ and modal volume
  $V \approx 0.8 (\lambda/n)^3$, along with its corresponding $H_z$
  and $E_y$ field profiles.}
\label{fig:fig2}
\end{figure}

\Figref{fig2} shows the peak spectral transfer $\eta_{\mathrm{max}} =
\Phi_\mathrm{max}/\Phi_\mathrm{BB}$, along with the normalized,
frequency-integrated heat transfer $H(\zeta)/H_{\max}(0)$ as a
function of $|\zeta|$, computed by integrating \eqref{cavity1}
numerically. Here $H_{\max}(0)$ denotes the maximum possible heat
transfer in the linear regime which occurs under the rate matching
condition $\gamma_e=\gamma_d$. (The inset shows a realistic structure
where such nonlinear radiation effects can potentially be observed.)
The largest increase in $\eta$ occurs when $\Delta T$ is largest and
so in the figure we consider the case $\Te = 0$, for multiple values
of $\Re \alpha / \Im \alpha$ and $\gamma_e/\gamma_d$. As $|\zeta|$
increases from zero, $\eta_\mathrm{max}$ increases and in certain
regimes becomes greater than one. At larger $\zeta$, the enhancement
is spoiled due to thermal broadening causing energy in the cavity to
leak out at a faster rate, thereby weakening nonlinearities and
causing $\eta_\mathrm{max} \to 0$ as $|\zeta| \to \infty$. The maximum
$\eta$ is determined by a competition between these two effects, with
thermal broadening becoming less detrimental and leading to larger
enhancements with decreasing $\gamma_e/\gamma_d$. We find that TPA
does not just enhance $\Phi(\omega)$ but also increases the total heat
transfer and in particular $\frac{H(\zeta)}{H_{\max}(0)} =
\frac{2\TeffNL}{\Delta T} \to \frac{2 \Td}{\Delta T} = 2$ in the limit
as $|\zeta| \to \infty$ (not shown), increasing monotonically with
increasing $\zeta$. As expected, $H$ is bounded by the largest
achievable effective temperature $\TeffNL \leq T_d$, or alternatively
by the maximum rate at which energy can be drawn from the phonon
bath. Examination of the reverse scenario ($\Te>\Td$), in which the
external bath is held at a higher temperature than the cavity, also
leads to similar enhancements. However, because only the internal bath
experiences nonlinear dissipation, the system exhibits non-reciprocal
behavior with respect to $\Td \rightleftharpoons \Te$, which is
evident in \eqref{Phi}. Moreover, we find that $\TeffNL$ in this
reverse scenario always decreases with increasing TPA. Such
non-reciprocity in the heat exchange is absent in linear systems and
could potentially be useful in technological applications, such as for
thermal rectification~\cite{Chang06,Otey10}.

To illustrate the range of thermal tunability offered by TPA, we
consider the on-resonance heat transfer $\Phi_\mathrm{max} \equiv
\Phi(\omega_0)$ in the highly non-equilibrium regime $\Te=0$ and for
$\Re \alpha =0$. While SPM offers some degree of tunability, we find
that TPA has a significantly larger impact on the radiation rate of
the cavity. In this regime, \eqref{Phi} simplifies and yields
$\Phi_{\mathrm{max}} = \frac{4\gamma_e\gamma_d}{\gamma^2} \left[ 1 +
  \frac{(-\Im\alpha)}{\gamma} ( 3 - 4\frac{\gamma_d}{\gamma}) \kB
  \Td\right]$, from which it follows that at small $\zeta \lesssim
0.5$, where \eqref{Phi} is applicable, $\Phi_\mathrm{max}$ scales
linearly with $\Td$ and depends on the ratio
$\frac{\gamma_d}{\gamma}$, increasing above its linear value whenever
$3\gamma_e \geq \gamma_d$. Furthermore, one finds that the largest
temperature variation, related to the slope
$\frac{(-\Im\alpha)}{\gamma} ( 3 - 4\frac{\gamma_d}{\gamma})$, is
obtained in the limit $\gamma_e \gg \gamma_d$ and $\gamma_e,\gamma_d
\to 0$, corresponding to a cavity with negligible linear emissivity,
large $\zeta \gg 1$ (for a fixed temperature and $\Im \alpha$), and
narrow bandwidth. Interestingly, we find that even for a finite $\zeta
\sim 0.4$ and $\gamma_e/\gamma_d \sim 20$, the emissivity of the
cavity can increase dramatically from $\Phi_\mathrm{max} \approx 0.2$
at $\Td=300$K to $\Phi_\mathrm{max}\approx 0.9$ at $\Td=600$K.

As mentioned above, it turns out that a nonlinear mechanical
oscillator interacting with a medium through nonlinear friction
exhibits similar spectral characteristics.~\cite{dykman75,dykman84}
However, in contrast to our photonic radiator, enhancements in the
spectral peak due to nonlinear friction are only observed when the
nonlinear dissipation rate is much larger than the linear loss rate,
or equivalently when the source of nonlinear friction is at a very
high temperature compared to the internal phonon
temperature.~\cite{dykman75} While realizing these experimental
conditions in mechanical oscillators, including the need to have
isolated linear and nonlinear heat baths operating at vastly different
temperatures, seems difficult, a photonic cavity offers alternative
ways of observing thermal radiation above the linear blackbody limit,
creating new opportunities for studying nonlinear damping. First,
while in the case of a mechanical oscillator one observes large
enhancements only when the internal dissipation and hence the
bandwidth $\gamma_d \to 0$, the introduction of an external radiative
channel in a photonic system enables large thermal enhancement with
finite $\gamma_d$ and hence larger bandwidths. In particular, as long
as the linear cavity losses are dominated by radiation to the external
bath, corresponding to the situation $\gamma_d \ll \gamma_e$, the
total cavity bandwidth $\gamma_e$ can be large while still allowing
internal losses to be dominated by nonlinear friction. Second, while
in the case of mechanical oscillators one observes nonlinear
enhancement only when the external (linear dissipation) temperature is
small compared to the internal (nonlinear dissipation) temperature,
$\Te \ll \Td$, interference effects associated with the presence of
the external bath in the photonic system ameliorate this
experimentally onerous constraint. In particular, the heat exchanged
between the photonic cavity and external bath depends on the sensitive
interference between reflected and emitted radiation from the cavity,
described by \eqref{cavity2}.  These interference effects result in
amplitude correlations $\sim \Im\alpha \langle s_{+}^* a a^*
\xi\rangle$, corresponding to the last term of \eqref{Phi}, whose
contribution to the heat transfer cannot be ignored in situations
where $T_e \lesssim T_d$. (In their absence, the spectrum of the
cavity resembles that of a mechanical oscillator and one can no longer
observe significant enhancements in thermal radiation unless $\Te \ll
\Td$.) For illustration, consider a situation in which the cavity and
external channel are held at temperatures $\Td=600$K and $\Te=300$K,
respectively. In this case, we find that the maximum transfer
increases from $\Phi_\mathrm{max}\approx 0.6$ at $\zeta=0$ to
$\Phi_\mathrm{max} \approx 1.2$ at $\zeta=1$ almost entirely due to
interference between the reflected and emitted radiation from the
cavity, in the absence of which $\Phi_\mathrm{max}$ actually decreases
with increasing $\zeta$.

\section{Conclusion}

We conclude by proposing a practical system where above mentioned
effects can potentially be observed. In order to reach the strongly
nonliner regime, it is desirable to have $|\zeta| = \frac{|\alpha| \kB
  T \gamma_e}{\gamma^2}\sim 1$. Given a choice of operating
temperature, the goal is therefore to design a cavity with a large
Purcell factor $\alpha/\gamma \sim Q/V$. If the goal is to observe
large enhancements from TPA, it is also desirable to operate with
materials and wavelengths where the nonlinear FOM $
\frac{n_2}{\lambda\beta_{\mathrm{TPA}}} \lesssim 1$, corresponding to
large TPA.~\cite{Boyd92} All of these conditions can be achieved in a
number of material systems and geometries. For illustration, we
consider the Ge nanobeam cavity shown on the inset of \figref{fig2}
and based on the family of nanobeam cavities described
in~\citeasnoun{Qimin11}, which supports a mode at
$\lambda=2.09\mu$m. At this wavelength, Ge has an index of refraction
$n\approx 4$ and Kerr coefficient $\chi^{(3)} \approx (1.2-11i)\times
10^{-17} (\mathrm{m}/\mathrm{V})^2$~\cite{Boyd92,Hon11}, corresponding
to a FOM $\approx 0.008$. This yields a mode with $\alpha \approx
0.001 (\chi^{(3)} / \varepsilon_0 \lambda^3)$, $Q \approx 10^8$, and
modal volume $V \approx 0.8 (\lambda/n)^3$, leading to $|\zeta|
\approx 1$ for operation at $T=1000$K. (Large Purcell factors such as
these were recently predicted in a similar, albeit silicon
platform~\cite{Qimin11}.)  We note that there are other possible
cavity designs, wavelength and material choices, including GaP and
ZnSe, and that it is also possible to operate with larger bandwidths
at the expense of larger temperatures and/or smaller mode
volumes. Because these thermal effects scale linearly with Purcell
factor, we believe that nanophotonic cavities with ultra-small modal
volumes and bandwidths are the most promising candidates for
experimental realization.  This is in contrast to the situation
encountered in traditional nonlinear devices involving incident
(non-thermal) light, where the threshold power for observing strong
nonlinear effects $\sim V/Q^2$ and therefore favors designs that
sacrifice modal volume in favor of smaller
bandwidths.~\cite{Bravo-Abad07}

Finally, we note that the predictions above offer only a glimpse of
the potentially interesting radiative phenomena that can arise in
passive nonlinear media at and out of equilibrium. In future work, it
may be interesting to consider the impact of other nonlinear phenomena
on thermal radiation, including free-carrier absorption and third
harmonic generation, as well as applications of the Kerr effect to
thermal rectification~\cite{Chang06,Otey10}.  In a related context of
optomechanics, the coupling of photonic and mechanical resonances
leads to novel nonlinear effects that are also manifested in the
radiation spectrum of photonic cavities, often studied in the presence
of incident, non-thermal radiation
pressure.~\cite{Kippenberg07,Aspelmeyer10} We believe that electronic
nonlinearities such as the Kerr effect in semiconductors offer an
alternative approach to exploring similar ideas involving nonlinear
fluctuations.

We are grateful to Mark Dykman for very helpful comments and
suggestions. This work was supported in part by the Army Research
Office through the Institute for Soldier Nanotechnologies under
Contract No. W911NF-13-D-0001, and by the National Science Foundation
under Grant No. DMR-145483.

\appendix

\section{Fokker--Planck Equation}
\label{app:FP}

In this appendix, we review the procedure for deriving the FP equation
[\eqref{dpdt}] from the corresponding nonlinear Langevin equation
[\eqref{cavity2}], which can be written in the following simplified
form:
\begin{align*}
  \dot{a}&=f(a,a^*)+D(a,a^*)\xi+\sqrt{2\gamma_e}s_{+} \\
  f(a,a^*)&=i(\omega_0-\alpha|a|^2)a-\gamma_a,
\end{align*}
where $D$ is the diffusion coefficient, and $\xi$ and $s_{+}$ are
delta-correlated white-noise sources obtained by taking the derivative
of standard Wiener processes~\cite{Higham01}, $\xi=\dot{W_\xi}$ and
$s_{+}=\dot{W_s}$. For a finite discretization time $\Delta t$, the
coupled-mode equations can be written as follows~\cite{MossBook07}:
\begin{multline}
  \frac{a(t)-a(t-\Delta t)}{\Delta t} = f(\lambda
  a(t)+(1-\lambda)a(t-\Delta t)) \\
  + D(\lambda a(t)+(1-\lambda)a(t-\Delta t))\frac{(W_1(t)-W_1(t-\Delta
    t))}
  {\Delta t} \\
  + \sqrt{2\gamma_e}\frac{(W_1(t)-W_1(t-\Delta t))}{\Delta t}
\end{multline}
where the choice of $0 \leq \lambda \leq 1$ determines the
corresponding Stochastic interpretation rule. Taylor expanding each
term and defining $\Delta a \equiv a(t)-a(t-\Delta t)$ and $\Delta W_i
= W_{i}(t)-W_{i}(t-\Delta t)$, with $\Delta W_i$ denoting a standard
Brownian increment with zero mean and variance $\langle \Delta W_i^*
\Delta W_i \rangle = k_B T \Delta t$, one finds the following
expression to $O(\Delta t)$,
\begin{multline}
  \Delta a = f(a,a^*)\Delta t + D\Delta W_1 + \lambda D\frac{\partial
    D}{\partial a}\Delta W_1 \Delta W_1 + \\
  \lambda D \frac{\partial D}{\partial a^*} \Delta W_1^* \Delta W_1 +
  \sqrt{2\gamma_e}\Delta W_2
  \label{eq:taylor}
\end{multline}
Transforming the Langevin equation into a FP PDE involves a standard
procedure ~\cite{MossBook07} and leads to an equation of the form
$\frac{\partial P}{\partial t} = -\sum_\alpha \frac{\partial}{\partial
  a_\alpha} K_\alpha P + \sum_{\alpha,\beta}
\frac{\partial^2}{\partial a_\alpha \partial a_\beta} K_{\alpha,\beta}
P$, where the FP coefficients are given by:
\begin{align}
  K_{\alpha}&=\lim_{\Delta t \to 0}\frac{\langle a_{\alpha}(t) -
    a_{\alpha}(t-\Delta t) \rangle}{\Delta t} \nonumber \\
  K_{\alpha,\beta}&=\lim_{\Delta t \to 0}\frac{\langle
    (a_{\alpha}(t)-a_{\alpha}(t-\Delta t))(a_{\beta}(t)-a_{\beta}
    (t-\Delta t)) \rangle}{\Delta t} \nonumber
\end{align}
Carrying out the above limiting procedures, one obtains the FP
equation given in \eqref{dpdt}.

\section{Perturbation Theory}
\label{app:pert}

In this appendix, we derive perturbative expressions for the energy
spectrum $\langle |a(\omega)|^2 \rangle$ and transfer function
$\Phi(\omega)$ of the nonlinear cavity. For convenience, we define
$\alpha = \alpha_1 - \alpha_2 i$, with $\alpha_2 = -\Im{\alpha} > 0$
as required by any passive nonlinear system.  We begin by defining a
perturbed cavity field $a(t) = a_0(t) + \da(t)$, where $a_0$ is the
linear cavity field and $\da$ is a correction of linear order in
$\alpha$. Plugging in the perturbed field into the coupled-mode
equations and ignoring terms $\mathcal{O}(\alpha^2)$ and higher, one
obtains the coupled equations:
\begin{align}
  \label{eq1}
  \dot{a}_0 &=(i\omega_{0}-\gamma)a_0 + \sqrt{2\gamma_{d}}\xi +
  \sqrt{2\gamma_{e}}s_{+} \\ \dot{\da} &=(i\omega_{0}-\gamma)\da-
  (i\alpha+\alpha_{2})|a_0|^{2}a_0 +
  \frac{\alpha_{2}}{\sqrt{2\gamma_{d}}}|a_0|^{2}\xi
\end{align}
Fourier transforming both equations, their solution to first order in
$\alpha$ can be written as:
\begin{align}
  \label{eq:a0}
  a_0(\omega)&= D(\omega)^{-1}\left[\sqrt{2\gamma_{d}}\xi(\omega) +
    \sqrt{2\gamma_{e}}s_{+}(\omega)\right] \\ \da(\omega) &=
  \frac{D(\omega)^{-1}}{\sqrt{2\gamma_{d}}}\mathcal{F}\left\{
    \alpha_{2}|a_0|^{2}\xi -
    (i\alpha_1+\alpha_{2})\sqrt{2\gamma_{d}}|a_0|^{2}a_0 \right\}
  \label{eq:a1}
\end{align}
where $D(\omega)\equiv i(\omega-\omega_{0})+\gamma$ and
$\mathcal{F} \equiv \int dt e^{-i\omega t}$ denotes the Fourier
transform operator.

\subsection{Energy spectrum}

We first compute the energy spectrum of the perturbed cavity which, to
first order in $\da$, is given by:
\begin{align}
  \langle |a(\omega)|^2 \rangle = \langle |a_0(\omega)|^{2} \rangle +
  2\Re\{ \langle a_0^{*}(\omega)\da(\omega) \rangle\},
\end{align}
As discussed below, the second term can be obtained by exploiting the
following linear, two-point correlation functions:
\begin{align}
  \label{eq:corr1}
  \langle a_0^{*}(\omega)\xi(\omega') \rangle &=
  \frac{\sqrt{2\gamma_{d}}\kB\Td\delta(\omega-\omega')}{D^{*}(\omega)}
  \\ \langle a_0(\omega)\xi(\omega') \rangle &=
  \frac{\sqrt{2\gamma_{d}}\kB\Td\delta(\omega+\omega')}{D(\omega)} \\
  \langle a^{*}_0(\omega)s_{+}(\omega') \rangle &=
  \frac{\sqrt{2\gamma_{e}}\kB\Te\delta(\omega-\omega')}{D^{*}(\omega)}\\
  \langle a_0(\omega)s_{+}(\omega') \rangle &=
  \frac{\sqrt{2\gamma_{e}}\kB\Te\delta(\omega+\omega')}{D(\omega)}\\
  \langle a_0^{*}(\omega)a_0(\omega') \rangle &= \frac{\gamma \kB\Teff
    \delta(\omega-\omega')}{D(\omega)D^{*}(\omega)},
  \label{eq:corr2}
\end{align}
where $\Teff = \frac{\gamma_e T_e + \gamma_d T_d}{\gamma}$ denotes the
linear effective temperature of the cavity. In deriving
\eqrefrange{corr1}{corr2}, we employed the fact that $s_{+}$ and $\xi$
are uncorrelated white-noise sources described by
\eqreftwo{sfdt}{xifdt}. \Eqref{corr2} is precisely the zeroth-order
term of the energy spectrum (in the absence of nonlinearities), while
the first-order correction is given by the more complicated
expression:
\begin{multline}
  \langle a_0^*(\omega) \delta a(\omega') \rangle \\ = \left\langle
    \frac{\left[\sqrt{2\gamma_d}\xi^*(\omega)+\sqrt{2\gamma_e}
        s_{+}^*(\omega)\right]\times \frac{1}{\sqrt{2\gamma_d}}
      G(\omega')}{D^*(\omega)D(\omega')}\right\rangle
  \label{eq:a0da}
\end{multline}
where the function
\begin{align*}
  &G(\omega) \equiv
  \mathcal{F}\left[\alpha_{2}|a_0|^{2}(\xi-\sqrt{2\gamma_{d}}a_0)-i\alpha_{1}\sqrt{2\gamma_{d}}|a_0|^{2}a_0
  \right]\nonumber \\
  &=\int_{-\infty}^{\infty}dx \int_{-\infty}^{\infty}dt
  e^{-i(\omega-x) t} \int_{-\infty}^{\infty} d\omega_{1}
  e^{-i\omega_{1}t} \int_{-\infty}^{\infty} d\omega_{2}
  e^{i\omega_{2}t} \\ &a_0^{*}(\omega_{1})a_0(\omega_{2}) \left[
    \alpha_2\left(\xi(x)-\sqrt{2\gamma_{d}}a_0(x)\right)
    -i\alpha_{1}\sqrt{2\gamma_{d}}a_0(x)\right],
\end{align*}
encapsulates the spectral response of the perturbed cavity field, here
simplified by exploiting the relation $\mathcal{F}\{|a_0|^2\} =
\mathcal{F}\{a_0\} \star \mathcal{F}\{a_0^*\}$. Focusing first on the
$\alpha_2$ terms of the numerator of \eqref{a0da}, one obtains:
\begin{multline*}
  \alpha_2 \int_{-\infty}^{\infty}dx \int_{-\infty}^{\infty}dt
  \int_{-\infty}^{\infty}d\omega_{1}
  \int_{-\infty}^{\infty}d\omega_{2}
  e^{-i(\omega'+\omega_{1}-\omega_{2}-x) t} \\
  a_0^{*}(\omega_{1})a_0(\omega_{2})\Big[\xi(x)-\sqrt{2\gamma_{d}}a_0(x)\Big]\Big[\xi^{*}(\omega)+\sqrt{\frac{\gamma_{e}}{\gamma_{d}}}s_{+}^{*}(\omega)\Big],
\end{multline*}
The ensemble average of the expression under the integrals involves
four-point correlation functions and is given by:
\begin{align*}
  \langle \ldots \rangle &= \langle
  a_0^{*}(\omega_{1})a_0(\omega_{2})\xi(x)\xi^{*}(\omega)\rangle \\ &+
  \sqrt{\frac{\gamma_{e}}{\gamma_{d}}} \langle
  a_0^{*}(\omega_{1})a_0(\omega_{2})\xi(x)s_{+}^{*}(\omega)\rangle \\
  &- \sqrt{2\gamma_{d}} \langle
  a_0^{*}(\omega_{1})a_0(\omega_{2})a_0(x)\xi^{*}(\omega)\rangle \\ &-
  \sqrt{2\gamma_{e}} \langle
  a_0^{*}(\omega_{1})a_0(\omega_{2})a_0(x)s_{+}^{*}(\omega)\rangle
  \label{eq:braket}
\end{align*}
Because the noise sources follow Gaussian distributions, four-point
correlation functions can be written in terms of products of two-point
correlation functions via Wick's theorem.~\cite{vankampen} Summing the
resulting two-point correlation functions, described by
\eqrefrange{corr1}{corr2}, one obtains the following three terms:
\begin{align*}
  &\Teff\left[\Td-\frac{2\gamma_{d}\Td+2\gamma_{e}\Te}{D(\omega)}\right]\frac{\delta(\omega-\omega')}{D^{*}(\omega)D(\omega')}, \\
  &\frac{2\gamma_{d}\Td+2\gamma_{e}\Te}{D(\omega)}\left[\Td-\Teff\right]\frac{\delta(\omega-\omega')}{D^{*}(\omega)D(\omega')}, \\
  &\frac{2\gamma_{d}\Td+2\gamma_{e}\Te}{D^{*}(-\omega)}\left[\Td-\Teff\frac{\gamma^{2}+i\omega_{0}\gamma}{\gamma^{2}+\omega_{0}^{2}}\right]\frac{\delta(\omega-\omega')}{D^{*}(\omega)D(\omega')}.
\end{align*}
It follows that the $\alpha_2$ term in \eqref{a0da} is given by:
\begin{multline}
  \frac{\alpha_{2}\kB^2}{|D(\omega)|^{2}}\left[\Teff \Td-\frac{2\gamma
      \Teff^{2}}{D(\omega)} +\frac{2\gamma \Teff
      \Td}{D(\omega)}- \frac{2\gamma \Teff^{2}}{D(\omega)} \right. \\
  \left.+\frac{\gamma \Teff \Td}{D^{*}(-\omega)} - \frac{2\gamma
      \Teff^{2}}{D(\omega)}\frac{\gamma^{2}+i\omega_{0}\gamma}{\gamma^{2}+\omega_{0}^{2}}
  \right]
\end{multline}
Note that the last two terms can be neglected since the quantities
$\propto D^{*}(-\omega)$ involve off-resonant, counter-rotating fields
and furthermore, our coupled-mode theory is only valid in the regime
$\gamma \ll \omega_{0}$. Performing a similar calculation for the
$\alpha_1$ term yields:
\begin{align}
  \frac{-\alpha_{1}\kB^2}{|D(\omega)|^{2}} \left[ \frac{2\gamma
      \Teff^{2}}{D(\omega)} + \frac{2\gamma \Teff^{2}}{D(\omega)} +
    \frac{2\gamma
      \Teff^{2}}{D^{*}(-\omega)}\frac{\gamma^{2}+i\omega_{0}\gamma}{\gamma^{2}+\omega_{0}^{2}}
  \right]
\end{align}
Putting together the above two expressions for both real and imaginary
$\alpha$ and neglecting counter-rotating terms, one obtains the energy
spectrum in \eqref{aneq}.

\subsection{Spectral transfer function}

The spectral transfer function is defined as the relative power
transfer from the cavity into the output channel divided by their
temperature difference,
\begin{align*}
  \Phi (\omega) = \frac{\langle |s_{-}(\omega)|^{2} \rangle - \langle
    |s_{+}(\omega)|^{2} \rangle}{\kB\Delta T}.
\end{align*}
To first order in $\alpha$, the outgoing power is given by:
\begin{multline}
  \langle |s_{-}(\omega)|^{2} \rangle = \langle
  |s_{+}(\omega)|^{2}\rangle + 2\gamma_{e} S_{aa}(\omega) \\ -
  2\sqrt{2\gamma_{e}}Re\{ \langle s_{+}^{*}(\omega)\left[a_0(\omega) +
    \da(\omega)\right]\rangle\},
\end{multline}
where the first and second terms are the incident power and energy
spectra of the cavity, obtained above, and so it only remains to
calculate the third term, or the interference between the incoming and
outgoing radiation. Following the same procedure as before, the
zeroth- and first-order correction terms are given by $ \langle
s_{+}^{*}(\omega) a_0(\omega) \rangle = \frac{\sqrt{2\gamma_e}
  T_d}{D^*(\omega)}$ and
\begin{multline*}
  \langle s_{+}^{*}(\omega)\delta a(\omega) \rangle =
  \frac{1}{\sqrt{2\gamma_{d}}} \int_{-\infty}^{\infty}dx
  \int_{-\infty}^{\infty}dt \int_{-\infty}^{\infty}d\omega_{1} \\
  \int_{-\infty}^{\infty}d\omega_{2}
  e^{-i(\omega'+\omega_{1}-\omega_{2}-x) t} \Big[
  \alpha_{2}a^{*}(\omega_{1})a(\omega_{2})(\xi(x)
  \\
  -\sqrt{2\gamma_{d}}a(x))s_{+}^{*}(\omega)-i\alpha_{1}\sqrt{2\gamma_{d}}a^{*}(\omega_{1})a(\omega_{2})a(x)s_{+}^{*}(\omega)
  \Big],
\end{multline*}
As before, these can be broken down into contributions from $\alpha_2$
and $\alpha_1$, which yields:
\begin{multline}
  \frac{\alpha_2 \sqrt{2\gamma_e} \kB^2}{|D(\omega)|^2} \Big\{ -\Teff
  \Te + \Te(\Td-\Teff) \\ + \frac{\Te D^*(\omega)}{D^{*}(-\omega)}
  \left[\Td -
    \Teff\left(\frac{\gamma^{2}+i\omega_{0}\gamma}{\gamma^{2}+\omega_{0}^{2}}\right)
  \right]\Big\},
\end{multline}
and
\begin{align}
  -\alpha_{1} \kB^2 \left[ \frac{2\sqrt{2\gamma_{e}}\Teff
      \Te}{D(\omega)^{2}} + \frac{\sqrt{2\gamma_{e}}\Teff
      \Te}{D(\omega)D^{*}(-\omega)} \right],
\end{align}
respectively. As before, the counter-rotating terms $\sim
D^{*}(\omega)$ can be neglected, leading to the following expression:
\begin{multline}
  \langle s_{+}^{*}(\omega)\delta a(\omega) \rangle = -
  4\alpha_{2}\gamma_{e}\Te\left[\Td - 2\Teff\right]\frac{\gamma^{2} \-
    (\omega-\omega_{0})^{2}}{|D(\omega)|^{4}} \\
  +8\alpha_{1}\gamma_{e}\Te\Teff\frac{\gamma
    (\omega-\omega_{0})}{|D(\omega)|^{4}}.
\end{multline}
Finally, after collecting like terms one obtains the spectral transfer
function in \eqref{Phi}.
 
\bibliographystyle{unsrt}
\bibliography{photon}

\begin{thebibliography}{10}

\bibitem{satoshi09}
Satoshi Kawata, Yashushi Inouye, and Prabhat Verma.
\newblock Plasmonics for near-field nano-imaging and superlensing.
\newblock {\em Nature Photonics}, 3:388--394, 2009.

\bibitem{lenert14}
Andrej Lenert, David~M. Bierman, Youngsuk Nam, Walker~R. Chan, Ivan Celanovic,
  Marin Soljacic, and Evelyn~N. Wang.
\newblock A nanophotonic solar thermophotovoltaic device.
\newblock {\em Nature Nanotechnology}, 9:126--130, 2014.

\bibitem{Rytov89}
S.~M. Rytov, V.~I. Tatarskii, and Yu.~A. Kravtsov.
\newblock {\em Principles of Statistical Radiophsics II: Correlation Theory of
  Random Processes}.
\newblock Springer-Verlag, 1989.

\bibitem{PolderVanHove71}
D.~Polder and M.~Van~Hove.
\newblock Theory of radiative heat transfer between closely spaced bodies.
\newblock {\em Phys. Rev.~B}, 4:3303--3314, 1971.

\bibitem{Eckhardt84}
W.~Eckhardt.
\newblock Macroscopic theory of electromagnetic fluctuations and stationary
  radiative heat transfer.
\newblock {\em Phys. Rev. A}, 29(4):1991--2003, 1984.

\bibitem{ReidRo12:review}
M.~T.~Homer Reid, Alejandro~W. Rodriguez, and Steven~G. Johnson.
\newblock Fluctuation-induced phenomena in nanoscale systems: harnessing the
  power of noise.
\newblock {\em Proc. IEEE}, 101(2):531--545, 2013.

\bibitem{Otey14:review}
Clayton~R. Otey, Linxiao Zhu, Sunil Sandu, and Shanhui Fan.
\newblock Fluctuational electrodynamics calculations of near-field heat
  transfer in non-planar geometries: A brief overview.
\newblock {\em J. Quan. Spect. Rad. Transfer}, 132:3--11, 2014.

\bibitem{Laroche06:prl}
M~Laroche, R~Carminati, and J.J. Greffet.
\newblock Coherent thermal antenna using a photonic crystal slab.
\newblock {\em Phys. Rev. Lett.}, 96:123903, 2006.

\bibitem{BasuZhang09}
S.~Basu, Z.~M. Zhang, and C.~J. Fu.
\newblock Review of near-field thermal radiation and its application to energy
  conversion.
\newblock {\em Int. J. Energy Res.}, 33(13):1203--1232, 2009.

\bibitem{Boyd92}
Robert~W. Boyd.
\newblock {\em Nonlinear Optics}.
\newblock Academic Press, California, 1992.

\bibitem{dykman75}
M.I. Dykman and M.A. Krivoglaz.
\newblock Spectral distribution of nonlinear oscillators with nonlinear
  friction due to a medium.
\newblock {\em Phys. Stat. Sol.(b)}, 68:111, 1975.

\bibitem{Bravo-Abad07}
Jorge Bravo-Abad, Alejandro~W. Rodriguez, Peter Bermel, Steven~G. Johnson,
  J.~D. Joannopoulos, and Marin Solja{\v{c}}i{\'{c}}.
\newblock Enhanced nonlinear optics in photonic-crystal nanocavities.
\newblock {\em Opt. Express}, 15(24):16161--16176, 2007.

\bibitem{Haus84}
H.~A. Haus.
\newblock {\em Waves and Fields in Optoelectronics}.
\newblock Prentice-Hall, Englewood Cliffs, NJ, 1984.

\bibitem{Hashemi09}
H.~Hashemi, A.~W. Rodriguez, J.~D. Joannopoulos, M.~Soljacic, and S.~G.
  Johnson.
\newblock Nonlinear harmonic generation and devices in doubly resonant kerr
  cavities.
\newblock {\em Phys. Rev.~A}, 79(1):013812, 2009.

\bibitem{berncallen}
William Bernard and Herbert~B. Callen.
\newblock Irreversible thermodynamics of nonlinear processes and noise in
  driven systems.
\newblock {\em Rev. Mod. Phys.}, 31:1017, 1959.

\bibitem{stratonovichBook93}
Rouslan~L. Stratonovich.
\newblock {\em Nonlinear Nonequilibrium Thermodynamics}.
\newblock Springer-Verlag, 1993.

\bibitem{klim}
Yu~L Klimontovich.
\newblock Nonlinear brownian motion.
\newblock {\em Physics-Uspekhi}, 37:737--767, 1994.

\bibitem{Kippenberg07}
P.~Del'Haye, A.~Schilesser, O.~Arcizet, T.~Wilken, R.~Holzwarth, and T.~J.
  Kippenberg.
\newblock Optical frequency comb generation from a monolithic microresonator.
\newblock {\em Nature}, 450:1214, 2007.

\bibitem{Zaitsev12}
Stav Zaitsev, Oleg Shtempluck, Eyal Buks, and Oded Gottlieb.
\newblock Nonlinear damping in a micromechanical oscillator.
\newblock {\em Nonlinear Dynamics}, 67:859--883, 2012.

\bibitem{Moser12}
Mark Dykman, editor.
\newblock {\em Fluctuating Nonlinear Oscillators: From Nanomechanics to Quantum
  Superconducting Circuits}, chapter~13.
\newblock Oxford Univrsity Press, 2012.

\bibitem{Cleland02}
A.N. Clelan and M.L. Roukes.
\newblock Noise processes in nanomechanical resonators.
\newblock {\em J.~Appl. Phys.}, 92(5):2758--2769, 2002.

\bibitem{Antonio12}
Dario Antonio, Damian~H. Zanette, and Daniel Lopez.
\newblock Frequency stabilization in nonlinear micromechanical oscillators.
\newblock {\em Nat. Comm.}, 3(806):1--6, 2012.

\bibitem{HoChan07}
H.B. Chan and C.~Stambaugh.
\newblock Fluctuation-enhanced frequency mixing in a nonlinear micromechanical
  oscillator.
\newblock {\em Phys. Rev.~B}, 73:224301, 2006.

\bibitem{Buks06}
R.~Almog, S.~Zaitsev, O.~Shtempluck, and E.~Buks.
\newblock High intermodulation gain in a micromechanical duffing resonator.
\newblock {\em Appl. Phys. Lett.}, 88:213509, 2006.

\bibitem{Andre12}
Stephan Andre, Lingzhen Guo, Vittorio Peano, Michael Mathaler, and Gerd Schon.
\newblock Emission spectrum of the driven nonlinear oscillator.
\newblock {\em Phys. Rev.~A}, 85:053825, 2012.

\bibitem{Kippenberg:07}
Tobias~J. Kippenberg and Kerry~J. Vahala.
\newblock Cavity opto-mechanics.
\newblock {\em Opt. Express}, 15(25):17172--17205, 2007.

\bibitem{Walls80}
P.D. Drummond and D.F. Walls.
\newblock Quantum theory of optical bistability
  $\uppercase\expandafter{\romannumeral 1\relax}$. nonlinear polarisability
  model.
\newblock {\em Journal of Physics A: Math. and Gen.}, 13:725, 1980.

\bibitem{dykman84}
M.I. Dykman and M.A. Krivoglaz.
\newblock Theory of nonlinear oscillators interacting with a medium.
\newblock {\em Soviet Scientific Reviews, A Phys.}, 5:265--442, 1984.

\bibitem{Luo04:thermal}
Chiyan Luo, A.~Narayanaswamy, G.~Ghen, and J.~D. Joannopoulos.
\newblock Thermal radiation from photonic crystals: A direct calculation.
\newblock {\em Phys. Rev. Lett.}, 93(21):213905, 2004.

\bibitem{Vahala03}
Kerry~J. Vahala.
\newblock Optical microcavities.
\newblock {\em Nature}, 424:839--846, 2003.

\bibitem{Qimin11}
Qimin Quan and M.~Loncar.
\newblock Deterministic design of wavelength scale, ultra-high q photonic
  crystal nanobeam cavities.
\newblock {\em Opt. Express}, 19:18529, 2011.

\bibitem{Painter10}
Amir~H. Safavi-Naeini, Thiago P.~Mayer Alegre, Martin Winger, and Osckar
  Painter.
\newblock Optomechanics in an ultrahigh-q two-dimensional photonic crystal
  cavity.
\newblock {\em Appl. Phys. Lett.}, 97:181106, 2010.

\bibitem{Johnson01:cavities}
Steven~G. Johnson, Attila Mekis, Shanhui Fan, and J.~D. Joannopoulos.
\newblock Molding the flow of light.
\newblock {\em Computing Sci. Eng.}, 3(6):38--47, 2001.

\bibitem{Rodriguez07:OE}
Alejandro Rodriguez, Marin Solja{\v{c}}i{\'{c}}, J.~D. Joannopulos, and
  Steven~G. Johnson.
\newblock $\chi^{(2)}$ and $\chi^{(3)}$ harmonic generation at a critical power
  in inhomogeneous doubly resonant cavities.
\newblock {\em Opt. Express}, 15(12):7303--7318, 2007.

\bibitem{vankampen}
N.G.Van Kampen.
\newblock {\em Stochastic Processes in Physics and Chemistry}.
\newblock North Holland, 3 edition, 2007.

\bibitem{Zwanzig01:book}
Robert Zwanzig.
\newblock {\em Nonequilibrium Statistical Mechanics}.
\newblock Oxford University Press, 2001.

\bibitem{MossBook07}
Frank Moss and P.V.E. McClintock.
\newblock {\em Noise in Nonlinear Dynamical Systems}, volume 1 and 2.
\newblock Cambridge University Press, 1 edition, 2007.

\bibitem{Haus84:coupled}
H.~A. Haus.
\newblock {\em Waves and Fields in Optoelectronics}.
\newblock Prentice-Hall, Englewood Cliffs, NJ, 1984.
\newblock Ch. 7.

\bibitem{Kardar07}
Mehran Kardar.
\newblock {\em Statistical Physics of Fields}.
\newblock Cambridge University Press, 2007.

\bibitem{Higham01}
Desmond~J. Higham.
\newblock An algorithmic introduction of numerical simulation of stochastic
  differential equations.
\newblock {\em SIAM Review}, 43:525--546, 2001.

\bibitem{Gieseler13}
Jan Gieseler, Lukas Novotny, and Romain Quidant.
\newblock Thermal nonlinearities in a nanomechanical oscillator.
\newblock {\em Nature Physics}, 9:806--810, 2013.

\bibitem{dykman92}
M.I. Dykman and P.V.E. McClintok.
\newblock Power spectra of noise-driven nonlinear systems and stochastic
  resonance.
\newblock {\em Physica D}, 58:10--30, 1992.

\bibitem{weiss87}
George~H. Weiss.
\newblock {\em Contemporary Problems in Statistical Physics}.
\newblock SIAM, 1 edition, 1987.

\bibitem{JoannopoulosJo08-book}
John~D. Joannopoulos, Steven~G. Johnson, Joshua~N. Winn, and Robert~D. Meade.
\newblock {\em Photonic Crystals: Molding the Flow of Light}.
\newblock Princeton University Press, second edition, February 2008.

\bibitem{Oono98}
Yoshitsugu Oono and Marko Paniconi.
\newblock Steady state thermodynamics.
\newblock {\em Prog. Theor. Phys. Suppl.}, 130:29--44, 1998.

\bibitem{Chang06}
C.~W. Chang, D.~Okawa, A.~Majumdar, and A.~Zettl.
\newblock Solid-state thermal rectifier.
\newblock {\em Science}, 314(5802):1121--1124, 2006.

\bibitem{Otey10}
Clayton~R. Otey, Wah~Tung Lau, and Shanhui Fan.
\newblock Thermal rectification through vacuum.
\newblock {\em Phys. Rev. Lett.}, 104(15):154301, 2010.

\bibitem{Hon11}
Nick~K. Hon, Richard Sored, and Bahram Jalali.
\newblock The third order nonlinear optical coefficients of $si$, $ge$ and
  $si_{1-x}ge_{x}$ in the midwave and longwave infrared.
\newblock {\em J. App. Phys.}, 110:011301--8, 2011.

\bibitem{Aspelmeyer10}
M.~Aspelmeyer, S.~Gr\"{o}blacher, K.~Hammerer, and N.~Kiesel.
\newblock Quantum optomechanics---throwing a glance.
\newblock {\em J. Opt. Soc. Am. B}, 27(6):A189--A197, 2010.

\end{thebibliography}

\end{document}